\documentclass{osa-article}

\journal{osajournal}

\articletype{Research Article}

\usepackage{lineno} 
\usepackage{graphicx}
\usepackage{subfigure}
\usepackage{amsmath}
\usepackage{bm}
\usepackage{enumerate}
\usepackage{algorithm}
\usepackage{algorithmic}
\usepackage{color}

\begin{document}
	
\title{Bayesian Recursive Information Optical Imaging: A Ghost Imaging Scheme Based on Bayesian Filtering}

\author{Long-Kun Du\authormark{1,3,5}, Chenyu Hu \authormark{2}, Shuang Liu \authormark{3,4}, Chenjin Deng\authormark{3,4}, Chaoran Wang\authormark{3,4}, Zunwang Bo\authormark{3,4}, Mingliang Chen\authormark{3,4}, Wei-Tao Liu\authormark{1,5,6,*}, Shensheng Han \authormark{2,3,4,*}}

\address{\authormark{1} Institute for Quantum Science and Technology, College of Science, National University of Defense Technology, Changsha 410073, China\\
	\authormark{2} Hangzhou Institute for Advanced Study, University of Chinese Academy of Sciences, Hangzhou 310024, China\\
	\authormark{3} Key Laboratory of Quantum Optics, Shanghai Institute of Optics and Fine Mechanics, Chinese Academy of Sciences, Shanghai 201800, China\\
	\authormark{4} Center of Materials Science and Optoelectronics Engineering, University of Chinese Academy of Sciences, Beijing 100049, China\\
	\authormark{5} Interdisciplinary Center of Quantum Information, National University of Defense Technology, Changsha 410073, China\\
	\authormark{6} Hunan Key Laboratory of Mechanism and Technology of Quantum Information, Changsha, Hunan, 410073, China}

\email{\authormark{*}Corresponding author: wtliu@nudt.edu.cn \\ 
	\authormark{*}Corresponding author: sshan@mail.shcnc.ac.cn} 

\begin{abstract}

Computational imaging~(CI) has been attracting a lot of interest in recent years for its superiority over traditional imaging in various applications. In CI systems, information is generally acquired in an encoded form and subsequently decoded via processing algorithms, which is quite in line with the information transmission mode of modern communication, and leads to emerging studies from the viewpoint of information optical imaging.
Currently, one of the most important issues to be theoretically studied for CI is to quantitatively evaluate the fundamental ability of information acquisition, which is essential for both objective performance assessment and efficient design of imaging system. 
In this paper, by incorporating the Bayesian filtering paradigm, we propose a framework for CI that enables quantitative evaluation and design of the imaging system, and demonstate it based on ghost imaging. In specific, this framework can provide a quantitative evaluation on the acquired information through Fisher information and Cramér-Rao Lower Bound (CRLB), and the intrinsic performance of the imaging system can be accessed in real-time.
With simulation and experiments, the framework is validated and compared with existing linear unbiased algorithms.
In particular, the image retrieval can reach the CRLB.
Furthermore, information-driven adaptive design for optimizing the information acquisition procedure is also achieved.
By quantitative describing and efficient designing, the proposed framework is expected to promote the practical applications of CI techniques.
\end{abstract}

\section{INTRODUCTION}

Computational imaging~(CI)~\cite{mait2018computational,sun20133d,barbastathis2019use} has emerged with much attention and developed as a new-type imaging paradigm.
Generally speaking, CI systems behave rather in line with the encoding-decoding information transmission mode adopted in Communication~\cite{blahut1987principles}. It typically involves well-designed modulation and acquisition schemes for optical signal via hardware, coupled with image retrieval algorithms through software.  
The analogy to communication systems allows CI systems to be understood and studied by incorporating the methodology of information theory, as the emerging of information optical imaging~\cite{shensheng2022review}.
Compared to traditional imaging which employs the ``direct point-to-point'' image information acquisition mode, CI is supposed to possess several further imaging capacities~\cite{hu2022ghost}, such as obtaining higher-dimensional information with lower-dimensional detectors by fully utilizing the imaging channel capacity~\cite{sun20133d}, increasing the 2-D spatial resolution by exploiting the discrepancy information of the high-dimensional light-field domain~\cite{tong2020spatial},
obtaining information at low sampling rates through compressive sensing~\cite{hu2019optimization},
and flexibly designing imaging systems according to specific tasks~\cite{ashok2008compressive}. 
These advantages have led to applications in various fields such as super-resolution imaging~\cite{gustafsson2000surpassing, li2019single, tong2021preconditioned}, imaging Lidar~\cite{gong2016three}, spectral imaging~\cite{liu2016spectral,sun2019single,yuan2021snapshot}, task-oriented imaging~\cite{liu2015edge,sun2019gradual,zhang2020image,zhang2022compressive,du2023information}, and so on.

Despite the above superiority, it remains crucial to establish a valid approach for the quantitative performance evaluation and the design of CI systems.  
The final imaging quality of CI systems depends not only on the encoding module, but also on the specific retrieval algorithm which may incorporate various kinds of prior knowledge~\cite{bian2018experimental,liu2020imaging}.
Therefore, relying solely on imaging quality as an indicator is inadequate for evaluating the information acquisition capability, and cannot effectively assist in the design of CI systems.
Since CI behaves rather analogous to communication systems, the ability of acquire information is supposed to be a reasonable criterion for system design~\cite{huck1988image}. 
However, extracting this intrinsic ability is challenging due to the varying results produced by different algorithms.
In scenarios where multiple measurements on different encoded information are conducted, the lack of an appropriate and timely performance evaluation makes it difficult to determine whether the acquired measurements are sufficient in real time. 
In addition, the pre-determined encoding light fields commonly used make it hard to perform adaptive imaging by using the idea of adaptive sensing~\cite{haupt2009adaptive,malloy2014near}, particularly in situations involving unknown moving objects, resulting in unnecessary resource consumption and limiting the CI system's flexibility to meet specific requirements.

Various information-theory-based studies in optical imaging~\cite{linfoot1955information,cox1986information,barrett1995objective,chao2016fisher} indicate that the information measure, such as information channel capacity and Fisher information, can be a powerful tool in quantitatively describing the imaging performance, including imaging accuracy and resolution~\cite{kosarev1990shannon,chao2016fisher}.
The design of imaging system can also be performed from an information-theoretic perspective~\cite{alter2000information}. 
However, most of these information-theoretic analyses are conducted after the entire detection process, which is unsuitable for dynamic imaging processes.
In this paper, we propose adopting Bayesian filtering as a processing framework that enables reliable and real-time evaluations of information acquisition for CI systems.
Bayesian filtering combines the state-evolution model and the state-measurement model to give consecutive estimation on the probability density of the state vector of a dynamic system~\cite{sarkka2013bayesian}. 
And the image information to be acquired in CI could be regarded as the state vector in Bayesian filtering~\cite{barrett2003foundations}.
From this perspective, the state-evolution and state-measurement models in Bayesian filtering are similar to the encoding and measurement process in the CI system. 
Specifically, 1) in general CI systems, the encoded information varies in different measurements, analogous to state-evolution; 2) the measurement of the encoded image information in CI aligns with the state-measurement in Bayesian filtering. 
By providing a probability density estimation of the state vector, Bayesian filtering facilitates a quantitative description of the intrinsic imaging performance in CI systems.

In this study, we demonstrate the proposed framework based on ghost imaging~(GI)~\cite{shapiro2012physics,shih2012physics,moreau2018ghost}, a typical CI system~\cite{hu2022ghost}.
From the information perspective, we show the efficacy of this framework on accessing the system's intrinsic information acquisition ability using the Fisher Information Matrix and Cram{\'e}r-Rao Lower Bound (CRLB) in image retrieval.
The numerical simulations demonstrate the capability to quantitatively estimate imaging performance and indicate the ability to reach this bound. And we compare the imaging performance with correlation-based linear estimation algorithms and validate the superiority of the proposed scheme.
Moreover, the framework realizes online performance bound estimation in a recursive way.
Additionally, we investigate the scalability of the proposed scheme by incorporating image priors, showcasing their effects on enhancing image retrieval, and demonstrate an information-driven adaptive encoding approach to optimize the information acquisition procedure in CI. 
Further, we demonstrate the effectiveness of the scheme through practical imaging experiments with quantitative estimation on the accuracy of the retrieved image.

\section{METHOD}

\subsection{Principle of ghost imaging}

\begin{figure}[htbp]
	\centering
	\includegraphics[width=0.6\linewidth]{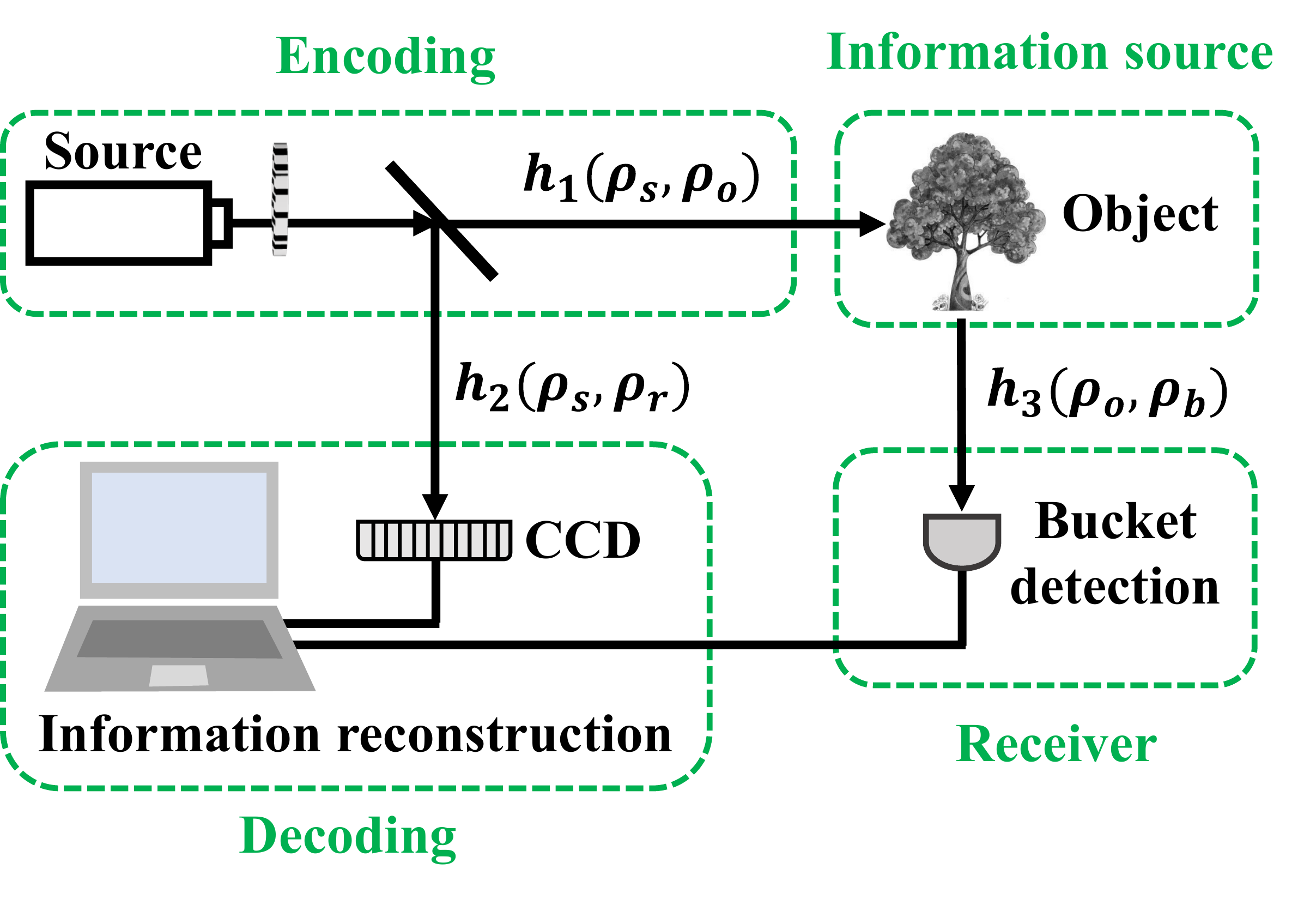}
	\caption{The schematic diagram of GI. The imaging process of GI is analogous to a communication system. The object is considered as the source in communication. Modulating the light source to illuminate the object can be understood as encoding information of the object. Bucket detection collects the intensity of the object's echo, serving as the receiver in communication. Finally, the information of the object is derived from the fluctuation correlation of the reference arm and bucket detection, serving as a decoding process.
	}
	\label{ghost_imaging}
\end{figure}

In GI, the object is modulated by a changing light field, and the information contained in the echo is recorded by bucket detection without spatial resolution.
By exploiting the fluctuation correlation between the illuminating field and the echo, the image of the object can be obtained.
A typical structure of GI and its analogy to the communication system is illustrated in Fig.~\ref{ghost_imaging}.
The light emitted from the source is divided into two arms by a beamsplitter. 
One is recorded by a detector with spatial resolution (CCD), known as the reference arm, while the other beam illuminates the object, referred to as the object arm. The echo of the object arm is collected by a bucket detector.

The light field that illuminates on the surface of object can be expressed as
\begin{equation}
I({\boldsymbol{\rho}_o}) = \int I ({\boldsymbol{\rho}_s}){h_1}({\boldsymbol{\rho}_s},{\boldsymbol{\rho}_o})d{\boldsymbol{\rho}_s},
\label{Eq_Ib}
\end{equation}
where $ I(\boldsymbol{\rho}_s) $ represents the light field at the source and ${h_1}({\boldsymbol{\rho}_s},{\boldsymbol{\rho}_o})$ represent the light intensity transfer functions from the light source to the object surface.
This article assumes that the illumination light fields are all pseudothermal \cite{erkmen2008unified}.
The light field interacts with the object and then propagates to the detector, which is collected by the bucket detection without spatial distribution.
The bucket detection only provides a total light intensity value, i.e.
\begin{equation}
    {I_b} = \int\int {I({\boldsymbol{\rho}_o})} t({\boldsymbol{\rho}_o}){h_3}({\boldsymbol{\rho}_o},{\boldsymbol{\rho}_b})d{\boldsymbol{\rho}_o}d{\boldsymbol{\rho}_b},
    \label{Eq_IB}  
\end{equation}
where $t({\boldsymbol{\rho}_o})$ represents the distribution of object's reflectance and ${h_3}({\boldsymbol{\rho}_o},{\boldsymbol{\rho}_b})$ represent the light intensity transfer functions from the object surface to the bucket detection surface.
The light intensity distribution of the reference arm $ {I}(\boldsymbol{\rho}_r) $ can be expressed as
\begin{equation}
	{I}(\boldsymbol{\rho}_r) = \int I (\boldsymbol{\rho}_s){h_2}(\boldsymbol{\rho}_s,\boldsymbol{\rho}_r)d{\boldsymbol{\rho}_s},
	\label{Eq_Ir}
\end{equation}
where  $h_2 (\boldsymbol{\rho}_s, \boldsymbol{\rho}_r) $ represents the intensity transfer function from the light source to the reference plane.
Experimentally setting ${h_1} ({\boldsymbol{\rho}_s}, {\boldsymbol{\rho}_o}) $ equals $h_2 (\boldsymbol{\rho}_s, \boldsymbol{\rho}_r) $, the light field of the reference arm $I({\boldsymbol{\rho}_r})$ is equal to the light field illuminated on the object $I({\boldsymbol{\rho}_o})$.
It is worth noting that in computational GI, the light field of the reference arm can be calculated, and the actual reference arm can be removed \cite{shapiro2008computational}. 
Therefore, the reflection (transmission) function of the object can be obtained by fluctuation correlation
\begin{equation}
	t({\boldsymbol{\rho}_o}) \propto \left\langle {\Delta I({\boldsymbol{\rho}_r})\Delta {I_b}}\right\rangle.  
	\label{fluctuation}
\end{equation}

The imaging process of GI is analogous to a communication system, as shown in Fig.~\ref{ghost_imaging}. The object is considered as the source in communication. Modulating the light source to illuminate the object can be understood as encoding information of the object. Bucket detection collects the intensity of the object's echo, serving as the receiver in communication. Finally, the fluctuation correlation is equivalent to a decoding process.
As a typical CI method, GI modulate the object's image information with controllable light fields and map it into lower dimensions, which can be described from the viewpoint of information optical imaging \cite{shensheng2022review,hu2022ghost}.

\subsection{Bayesian filtering}

Bayesian filtering is a recursive estimation method to give consecutive probability estimation on the state of a dynamic system based on Bayes' theorem. The Bayes' theorem can be expressed as  
\begin{equation}
	p(\boldsymbol{x}|\boldsymbol{z}) =\frac{p(\boldsymbol{z}|\boldsymbol{x}) p(\boldsymbol{x})}{p(\boldsymbol{z})}.
\end{equation}
where $p(\boldsymbol{x})$ and $p(\boldsymbol{z})$ are the probabilities of observing $\boldsymbol{x}$ and $\boldsymbol{z}$ respectively without any given conditions, which are known as the prior probability, and $ p(\boldsymbol{z})\neq 0 $.
$ p(\boldsymbol{x}|\boldsymbol{z}) $ is a conditional probability of event $ \boldsymbol{x} $ occurring given $ \boldsymbol{z} $, which is called the posterior probability. 

In the Bayesian filtering paradigm, a dynamic system is constantly evolving with time. By measuring at different time, the system state is recursively estimated from those measurements. Specifically, for a dynamic system, the state is described by the vector $ \boldsymbol{x}_k $ in $R^{n_x} $, where $ k $ represents the time coordinate and $R^{n_x} $ stands for ${n_x}$-dimensional space. The state-evolution is described by a First-order Markov process
\begin{equation}
	\boldsymbol{x}_k = f_{k-1} (\boldsymbol{x}_{k-1}, {w}_{k-1}), 
	\label{state_evolution}
\end{equation}
where $f_{k-1} $ represents state transition relation of $\boldsymbol{x}$, and ${w}_{k-1} $ is referred to as process noise sequence.
On the other hand, the state-measurement values $\boldsymbol{z}_k$ are related to the system state via the measurement equation
\begin{equation}
	\boldsymbol{z}_k = h_k (\boldsymbol{x}_k, {v}_k),
	\label{measurement}
\end{equation}
where $ h_k $ represents the process of measurement and $ {v}_k $ is referred to as the measurement noise sequence.
In the process, $ \boldsymbol{x}_k $ is estimated based on the sequence of all available $\boldsymbol{Z}_k \triangleq \{z_i, i = 1,\cdots, k\}$ up to time $k$, and the degree of belief in state $ \boldsymbol{x}_k $ is recursively quantified. 
The Bayesian filtering paradigm employs a two-step approach of "prediction" and "update", to recursively estimate the probability distribution of $ \boldsymbol{x}_k $ from the previous object state vector $ \boldsymbol{x}_{k-1}$.
Suppose that the probability distribution $p (\boldsymbol{x}_{k-1} | \boldsymbol{Z}_{k-1}) $ has been obtained after $ k-1 $ measurement $ \boldsymbol{Z}_{k-1} $, the two-step process is as follows:

1.Prediction. According to the system state-evolution model, a probability density of $ \boldsymbol{x}_k $ at time $k$ can be predicted, without the $k-$th measurement
\begin{equation}
	p\left(\boldsymbol{x}_k \mid \boldsymbol{Z}_{k-1}\right)=\int p\left(\boldsymbol{x}_k \mid \boldsymbol{x}_{k-1}\right) p\left(\boldsymbol{x}_{k-1} \mid \boldsymbol{Z}_{k-1}\right) \mathrm{d} \boldsymbol{x}_{k-1},
	\label{Eq_prediction}
\end{equation}
where $ p (\boldsymbol{x}_k|\boldsymbol{x}_{k-1}) $ can be determined by the state-evolution model and the known probability density function of ${w}_{k-1}$.

2. Update. At time $k$ with the measurement $ \boldsymbol{z}_k $ available, the update stage is executed using Bayes' theorem
\begin{equation}
	\begin{aligned} 
		p\left(\boldsymbol{x}_k \mid \boldsymbol{Z}_k\right) & =p\left(\boldsymbol{x}_k \mid \boldsymbol{z}_k, \boldsymbol{Z}_{k-1}\right) \\ 
		& =\frac{p\left(\boldsymbol{z}_k \mid \boldsymbol{x}_k, \boldsymbol{Z}_{k-1}\right) p\left(\boldsymbol{x}_k \mid \boldsymbol{Z}_{k-1}\right)}{p\left(\boldsymbol{z}_k \mid \boldsymbol{Z}_{k-1}\right)} \\ 
		& =\frac{p\left(\boldsymbol{z}_k \mid \boldsymbol{x}_k\right) p\left(\boldsymbol{x}_k \mid \boldsymbol{Z}_{k-1}\right)}{p\left(\boldsymbol{z}_k \mid \boldsymbol{Z}_{k-1}\right)}
	\end{aligned}
	\label{Eq_update}
\end{equation}
where $ p\left(\boldsymbol{z}_k \mid \boldsymbol{Z}_{k-1}\right)=\int  p\left(\boldsymbol{z}_k \mid \boldsymbol{x}_{k}\right) p\left(\boldsymbol{x}_k \mid \boldsymbol{Z}_{k-1}\right) d\boldsymbol{x}_k $, and $ p(\boldsymbol{z}_k|\boldsymbol{x}_k) $ is jointly determined by the state measurement model and the known probability density function of $ {v}_{k-1} $.
In addition, $ p(\boldsymbol{x}_0|\boldsymbol{z}_0)\triangleq p(\boldsymbol{x}_0) $, representing prior information about the system state before any observations are made.

The recurrence relations in Eq.~(\ref{Eq_prediction}) and Eq.~(\ref{Eq_update}) serve as the foundation for the optimal Bayesian solution. The knowledge of the posterior density $p\left(\boldsymbol{x}_k \mid \boldsymbol{Z}_k\right)$ enables the optimal state estimate, and the measurement accuracy of state estimate (e.g., covariance) can also be derived from $p\left(\boldsymbol{x}_k \mid \boldsymbol{Z}_k\right)$.

\subsection{Quantitative ghost imaging with Bayesian filtering}

GI utilizes low-dimensional information to gradually recover high-dimensional information, which can be modeled as a dynamic process.
For the object, the evolution of states over time can be expressed as $t_k = A_{k-1} t_{k-1}+{w}_{k-1}$, which is a linear process. $ {t}_k $ is a vector of dimension $n_x$ that describes the state, $A_{k-1}$ is a known matrix of dimension $n_x \times n_x $ defining the object evolution functions and $ {w}_{k-1} $ represents process noise.
As Fig.~\ref{ghost_imaging}, the interaction between the light field and the object during the $(k-1)$-th illumination resulted in $ {s_{k-1}}({\boldsymbol{\rho}_o}) = {I_{k-1}}({\boldsymbol{\rho}_o}){t_{k - 1}}({\boldsymbol{\rho}_o}) $, where $I_{k-1}$ is a known matrix of dimension $n_x \times n_x $ determined by the modulated light field. Then the interaction during the $k$-th illumination can be expressed as $ {s_k}=I_k t_k= B_k s_{k-1} + {I_k} {{w}_{k-1}} $, where ${B_k}={I_k} I_{k-1}^{-1} A_{k-1}$. 
Therefore, the state-evolution process and state-measurement process can be summarized as 
\begin{equation}
	\begin{aligned}
	t_k &= A_{k-1} t_{k-1}+{w}_{k-1},\\
	{s_k} &=  B_k s_{k-1} + {I_k} {{w}_{k-1}},\\
	{d_k} &= {H_k}{s_k} + {{v}_k}.
	\end{aligned}
 \label{iterative_equation}
\end{equation}
where $ H_k $ (of dimension $ 1 \times n_x $) is the measurement matrix of detection, representing the collection of signals, and $ {v_k} $ represents measurement noise.
Noteworthy, $ I_k({\boldsymbol{\rho}_o}) $ is the result of discretization of the light field in Eq.~(\ref{Eq_Ib}).
In ghost imaging, the noise in bucket detection is often regarded as Gaussian white noise, and the noise in state-evolution can also be assumed Gaussian. They can be respectively represented as $ p({w}) \sim N(0,Q) $ with variance $Q$, and $ p({v}) \sim N(0,R) $ with variance $R$, where $Q$ and $R$ are the Covariance matrix of the noise.
As the process described above is linear, the equation in Eq.~(\ref{Eq_prediction}) and Eq.~(\ref{Eq_update}) can reduce to the Kalman filter. 
This provides an optimal estimation solution, given the assumption of linearity.

The specific structure of Bayesian recursive ghost imaging is presented in the $\textbf{Algorithm}$ table.
\begin{algorithm}[h]
	\renewcommand{\thealgorithm}{}
	\caption{Bayesian recursive ghost imaging.}
	\label{alg_1}
	\begin{algorithmic}[1]
		\REQUIRE Initial state $t_0$, initial covariance $P_0$, state transition matrix $A$, observation matrix $H$, process noise covariance $Q$, measurement noise covariance $R$
		\ENSURE Estimated state $\hat{t}$ and estimated covariance $P$
		\FOR{$k=1$ to $N$}
		\STATE Prediction:
          \STATE $\hat{t}_k^{\prime} = A_{k-1} \hat{t}_{k-1} $
          \STATE $\hat{s}_k^{\prime} = B_k \hat{s}_{k-1} $
		  \STATE $P_k^{\prime} =  A_{k-1} P_{k-1} A_{k-1}^T + Q$
		\STATE Update:
		  \STATE $K_k = P_k^{\prime}(H_{k}I_{k})^T(H_{k}I_{k} P_k^{\prime} (H_{k}I_{k})^T + R)^{-1}$
		  \STATE $\hat{t}_k = \hat{t}_k^{\prime} + K_k(d_k - H_{k}\hat{s}_k^{\prime})$
		  \STATE $\hat{s}_k = I_k \hat{t}_{k} $
		  \STATE $P_k = (I - K_k H_{k}I_{k})P_k^{\prime}$
		\RETURN $ \hat{t}_k $ and $ P_k $
		\ENDFOR
	\end{algorithmic}
\end{algorithm}
As previously introduced, the iterative process of the algorithm is divided into two stages: Prediction and Update. 
In the $\textbf{Algorithm}$, $ P_k^{\prime} = E[( \hat{t}_k^{\prime} - {t_k})(\hat{t}_k^{\prime} - {t_k})^T] $ represents the error covariance matrix between the predicted value and the real value, and $ {P_k} = E[( \hat {t}_k - {t_k})(\hat{t}_k - {t_k} )^T] $ represents the error covariance matrix between the estimated value and the real value.
In this process, we can timely obtain the state estimation result $\hat{t}_k$ and the corresponding covariance matrix $ P_k $.
$k$ is the update step and $N$ is the total number of measurements.
The initial values of $t_0$ and $P_0$ can be set based on prior experience, and even with some deviation, the filtering results will gradually approach the true value. 
If the initial value is unbiased, the entire filtering process is unbiased.

In the estimation process, we can compute the Fisher information of object estimation and derive CRLB based on the Fisher information.
In mathematical statistics, the Fisher information is defined as
\begin{equation}
	\mathbb{J}(t)=\mathbb{E} \left[-\frac{\partial^2 \ln p(\boldsymbol{d} ; t)}{\partial t^2}\right],
\end{equation}
which is a way of measuring the amount of information that an observable random variable $\boldsymbol{d}$ carries about an unknown parameter $t$ upon which the probability of $\boldsymbol{d}$ depends. 
As for CRLB, it states that the variance of an unbiased estimator cannot be lower than the reciprocal of the Fisher information, i.e, 	$\mathbb{V}(t) \geq \mathbb{J}_k^{-1}$.
CRLB can also be used to bound the variance of biased estimators of given bias \cite{cramer1999mathematical,rao1992information}.
We can estimate the CRLB of images during sequential iteration as 
\begin{equation}
	\rm{CRLB}_k=\mathbb{J}_k^{-1},
\end{equation}
representing the lower bound of the estimation error.
For estimation of multiple parameters, the Fisher information matrix is
\begin{equation}
	[\mathbb{J}(\boldsymbol{t})]_{i j}=\mathbb{E}\left[-\frac{\partial^2 \ln p(\boldsymbol{d} ; \boldsymbol{t})}{\partial \boldsymbol{t}_i \partial \boldsymbol{t}_j}\right].
\end{equation}
Then each parameter has its own lower bound on the variance $\mathbb{V}\left(\boldsymbol{t}_i\right) \geq {\rm CRLB}_{ii}=\left[\mathbb{J}^{-1}(\boldsymbol{t})\right]_{ii}$ when being estimated. 
The CRLB consequently transforms into a matrix from Fisher information with multiple-parameters.
Combining Bayesian dynamic filtering in image reconstruction of GI, the lowest limit of image estimation accuracy can be expressed as
\begin{equation}
	{P_{k}} \buildrel \Delta \over = \mathbb{E}\{ ({\hat{t}_{k}} - {t_k}){({\hat{t}_{k}} - {t_k})^T}\}  \ge \mathbb{J}_k^{-1}
\end{equation}
where ${\mathbb{J}_k} $ is the Fisher information matrix.

In our method, the iterative relationship for the Fisher information can be derived as
\begin{equation}
	\mathbb{J}_{k+1} = (Q_k+ A_{k} \mathbb{J}_{k} A_{k}^{T})^{-1} + (H_{k+1}I_{k+1})^{T} R_{k+1}^{-1} H_{k+1}I_{k+1},
\end{equation}
which is updated in real-time through recursive calculations based on each measurement. The Fisher information matrix of the initial state is
\begin{equation}
	\begin{aligned}
		\mathbb{J}_0 &=  \mathbb{E} \{ P_0^{-1}(\hat t_{0}- {t}_{0})(\hat t_{0}- {t}_{0})^T [P_0^{-1}]^T\}\\
		&=P_0^{-1} P_0 P_0^{-1}=P_0^{-1},
	\end{aligned}
\end{equation}
where $P_0$ represents the prior information on the initial covariance matrix of the object.
With the above modeling, the estimation accuracy can be evaluated by Fisher information and CRLB during the imaging process.

\section{SIMULATION RESULTS} 

In this section, we employ numerical simulations to validate our theoretical framework.
First, we define the signal-to-noise ratio of detection.
GI utilizes the fluctuation correlation of the light field to extract image information, where only the signal fluctuation is relevant. In this context, the signal-to-noise ratio of bucket detection (BSNR) is defined as the ratio of the signal variance to the noise variance \cite{li2021enhancing}
\begin{equation}
	{\rm BSNR}= 10\log _{10} \frac{\left\langle I_b^2\right\rangle-\left\langle I_b\right\rangle^2}{{\sigma_n^2}},
\end{equation}
which is measured in dB, with ${\sigma_n^2}$ representing the variance of noise in bucket detection.

\subsection{Quantitative Description and Validation of Bayesian Ghost Imaging}

In simulation, pseudothermal light field with a negative exponential distribution is generated to illuminate the object, as Fig.~\ref{ghost_imaging}.
The speckle size is the same as the pixel size of the CCD in the reference arm, which is also the resolution of the final image.
After reflection by the object, a bucket detector collects the reflected light field to produce bucket detection value.
In this process, Gaussian noise is added to simulate detection noise.
The imaging process is modeled as a dynamic process according to Eq.~(\ref{iterative_equation}), and incorporated into Bayesian filtering algorithm.
Based on the model of previous section, the initial distribution of each point in the image can be regarded as a Gaussian, and its probability density function can be expressed as $ {N}(t_0(i,j),P_0(i,j)) $. The joint probability density function of the entire image can be expressed as
\begin{equation}
	\begin{aligned}
	 p[t_0(\boldsymbol{\rho})]&=  {N}({t_0(i,j)},{P_0(i,j)}), \\
	 t_0  &=\alpha \boldsymbol{e}, P_0 =\beta E,
	\end{aligned}
\label{Eq_p0}
\end{equation}
where $t_0$ and $P_0$ are the initial values, $ \boldsymbol{e} $ represents the unit vector with dimension $n_x$, and $E=diag(1,1,...,1)$ represents the unit matrix with dimension $n_x \times n_x $.
The uniform vector of initial value $t_0$ represents the initial state in the absence of spatial distribution information.
The initial covariance matrix $P_0$ is diagonal, indicating that the data points in the initial vector are statistically independent.

\begin{figure}[htbp]
	\centering
	\includegraphics[width=\linewidth]{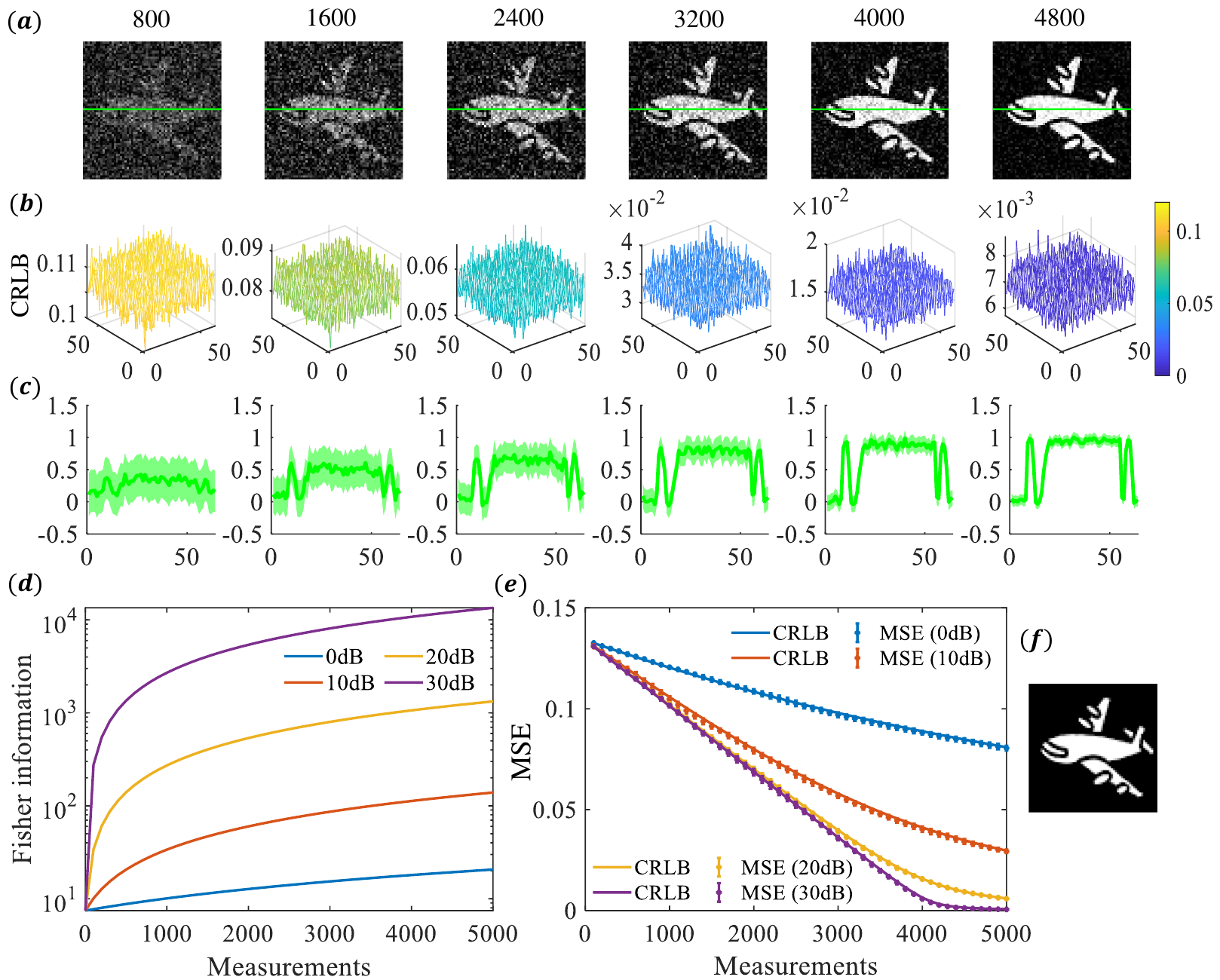}
	\caption{Imaging results and quantitative description of information acquisition in BGI. (a) The imaging results of BGI under different measurements, with BSNR is 20dB. (b) The corresponding  CRLB of each point in (a). (c) The estimated standard deviation of a one-dimensional cross-section in (a) obtained from CRLB. (d) The average Fisher information of the image varies with the number of measurements under different BSNR. (e) The variation of CRLB with the number of measurement under different BSNR in BGI and the corresponding MSE of imaging results. (f) The real object.}
	\label{airplane_1}
\end{figure}

The results of Bayesian ghost imaging (BGI) and their quantitative error descriptions are presented in Fig.~\ref{airplane_1}.
In GI, for unknown objects, the mean and variance of objects can be estimated through bucket detection signals (see details in the Appendix).
Then, in this case, the initial result of the object can be set as $ \alpha = 0.181 $ and $\beta = 0.134 $.
It is worth noting that despite a certain error in the initial value setting, the estimated results will gradually converge towards the true value as the number of measurements increases, leading to a reduction in the error caused by the initial value setting.
Fig.~\ref{airplane_1} (a) illustrates the BGI results of different frame numbers, ranging from 800 to 4800, displayed from left to right, under a BSNR of 20dB.
The ground truth is depicted in Fig.~\ref{airplane_1} (f), with an image resolution of $n_x=64*64$. The image are within the range of [0-1], and the grayscale value is 10.
The Fisher information and CRLB for the image can be acquired during the estimation process. Specifically, the Fisher information and CRLB are represented by an $n_x \times n_x $ matrix, where the diagonal elements correspond to the information content of each pixel.
Fig.~\ref{airplane_1}(b) represents the corresponding CRLB for each point in Fig.~\ref{airplane_1}(a).
Since all points on the object are equivalent, the decrease in CRLB at each point is essentially identical.
In Fig.~\ref{airplane_1}(b), the CRLB reaches the order of $ 10^{-3} $ at 4800 frames, indicating the result is close to the ground truth. 
The variation of the average Fisher information of the entire image is shown by the yellow curve in Fig.~\ref{airplane_1}(d), representing the change in Fisher information with the number of measurements under BSNR of 20dB.
As the measurement increases, the Fisher information gradually increases, and conversely, the corresponding CRLB gradually decreases and approaches zero.
To more intuitively represent the estimation errors scale of the images, Fig.~\ref{airplane_1}(c) provides the estimated standard deviation of the one-dimensional cross-section in Fig.~\ref{airplane_1}(a). 
This standard deviation is obtained from the square root of CRLB, and the light green area clearly indicates how the estimated  error varies with the number of measurements.
The impact of signal-to-noise ratio on the Fisher information are illustrated in Fig.~\ref{airplane_1}(d). 
The blue, red, yellow, and purple lines represent the variation of mean Fisher information with the number of measurements at BSNR of 0 dB, 10 dB, 20 dB, and 30 dB, respectively.
Meanwhile, Fig.~\ref{airplane_1}(e) represents the variation of CRLB with measurement times under different signal-to-noise ratios. The solid blue, red, yellow, and purple lines represent the variation of CRLB under BSNR of 0 dB, 10 dB, 20 dB, and 30 dB, respectively.
Specifically, higher BSNR values lead to a more rapid increase in Fisher information and decrease in the CRLB.
To verify the CRLB results, we also evaluated the mean square error (MSE) of the actual imaging results
\begin{equation}
	\mathrm{MSE} =\frac{1}{H \times W} \sum_{i=1}^H \sum_{j=1}^W(t(i, j)-\hat{{t}}(i, j))^2,
	\label{Eq_mse}
\end{equation}
where $ H $ and $ W $ represent the height and width of the image.
The asterisk `*' in Fig.~\ref{airplane_1}(e) represents the average of 10 sets of simulated MSE results, with blue, red, yellow, and purple representing BSNR of 0 dB, 10 dB, 20 dB, and 30 dB, respectively.
The results for each point are calculated using 10 sets of data, and the corresponding error bar is also provided.
As can be seen, the comparison of the actual imaging results' MSE with our predicted CRLB indicates a good match, which confirms the effectiveness of our predicted CRLB results.
Furthermore, it also indicates that this method can achieve the theoretical information error bound.

\begin{figure}[htbp]
	\centering
	\includegraphics[width=\linewidth]{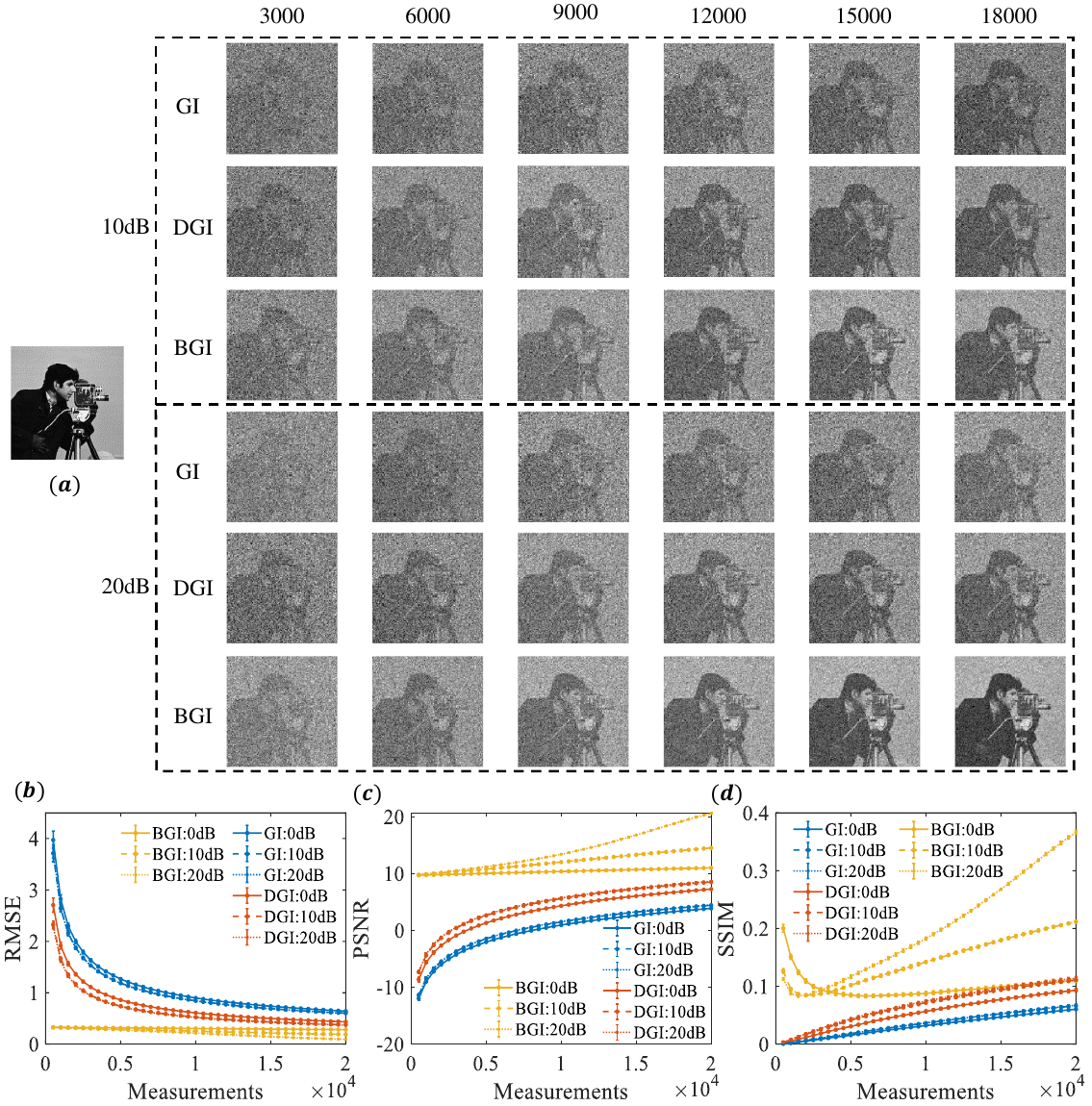}
	\caption{Imaging results of BGI, compared with GI and DGI. (a) The target cameraman of imaging. The upper three rows on the right side of figure (a) represent the imaging results of GI, DGI and BGI with BSNR is 10dB, and the lower three rows represent the result of GI, DGI and BGI with BSNR is 20dB. (b), (c), and (d) depict the RMSE, PSNR, and SSIM of the imaging results for GI, DGI and BGI, with BSNR of 0dB, 10dB, 20dB, respectively.}
	\label{cameraman_1}
\end{figure}
Fig.~\ref{cameraman_1} compares the imaging quality of BGI with traditional ghost imaging. Since that BGI is a linear algorithm, it is compared with the ghost imaging (GI) in Eq.~(\ref{fluctuation}) \cite{chan2010optimization} and differential ghost imaging (DGI) \cite{ferri2010differential}, which are linear as well.
The object for imaging is shown in Fig.~\ref{cameraman_1}(a), with a resolution of 128*128 pixels, and a grayscale of 256 levels (8-bit).
The figures on the right side of Fig.~\ref{cameraman_1}(a) represents the imaging results of GI, DGI, and BGI, respectively.
The upper three rows are the imaging results of BSNR with 10dB, and the lower three rows are the imaging results of BSNR with 20dB.
The final imaging results of BGI exhibit a significant improvement compared to GI and DGI.
To facilitate a clearer comparison, we present the imaging results of BGI, GI, and DGI in Fig.~\ref{cameraman_1} (b), (c), and (d), utilizing Root Mean Square Error (RMSE), Peak Signal-to-Noise Ratio (PSNR), and Structural Similarity Index (SSIM) as evaluation metrics for assessing the quality of the images, respectively, as Eq.~(\ref{Eq_rmse})
\begin{subequations}
	\begin{align}
	&\mathrm{RMSE} =\sqrt{MSE},\\
	&\mathrm{PSNR} =10 \log _{10}\left(\frac{\left(2^n-1\right)^2}{MSE}\right),\\
	&\mathrm{SSIM}=\frac{\left(2 \mu_t \mu_{\hat{t}}+C_1\right)\left(2 \sigma_{t {\hat{t}}}+C_2\right)}{\left(\mu_t^2+\mu_{\hat{t}}^2+C_1\right)\left(\sigma_t^2+\sigma_{\hat{t}}^2+C_2\right)}.
	\end{align}
	\label{Eq_rmse}
\end{subequations}
In Eq.~(\ref{Eq_rmse}b), $n$ represents the bit depth of the image.
In Eq.~(\ref{Eq_rmse}c), SSIM uses the basic default form, where $\mu$, $\sigma$ and $\sigma_{t \hat{t}}$ represent the mean, standard deviation, and cross covariance, respectively.
$C1=(0.01 * L)^ 2$, $C2=(0.03 * L)^ 2$, $C3=C2/2$, and $ L $ is the specified DynamicRange value.
In Fig.~\ref{cameraman_1} (b), (c), and (d), the imaging results of BGI, GI, and DGI are represented by yellow, blue, and red curves, respectively. The solid, dashed, and dotted lines indicate the imaging results with the BSNR at 0 dB, 10 dB, and 20 dB, respectively. Each data point is the statistical result of 10 sets of data, and the corresponding errorbar is indicated in the figure.
Obviously, the number of measurements increasing leads to a decrease in the RMSE, and an increase in PSNR and SSIM,  indicating that the imaging quality is gradually improving.
And based on these parameters, BGI demonstrates higher imaging quality compared to GI and DGI.
With an increase in BSNR, BGI shows a significant improvement, whereas DGI and GI exhibit less enhancement.
In Fig.~\ref{cameraman_1} (d), the SSIM of BGI first decreasing and then increasing. The possible reason is that SSIM is closely related to the overall level of the image, and the initial measurement may affect the overall image level, resulting in a larger initial SSIM.
These results suggest that BGI can improve the imaging quality.

\subsection{Incorporating Prior Information for Imaging}

\begin{figure}[htbp]
	\centering
	\includegraphics[width=0.9\linewidth]{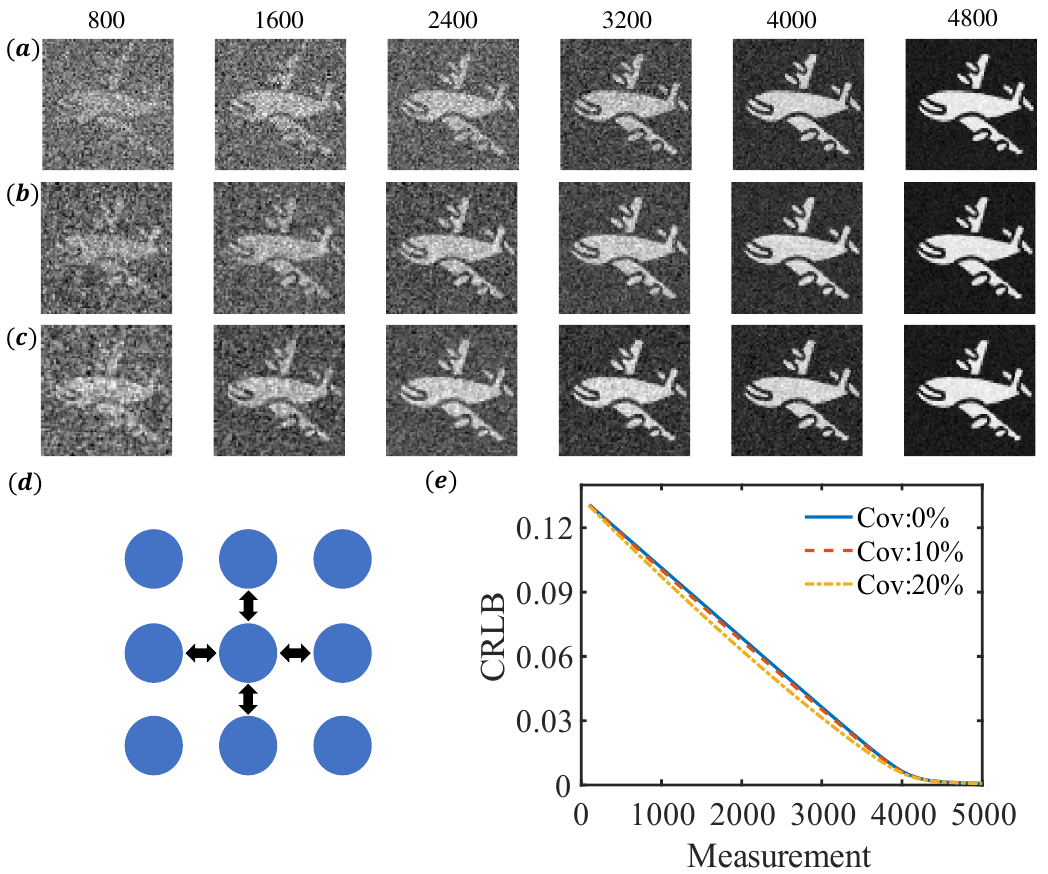}
	\caption{BGI results with prior information of spatial constraints. (a), (b), and (c) represent the imaging results under covariance matrix constraints $ \mathrm{Cov}:0\% $, $ \mathrm{Cov}:10\%$, and $ \mathrm{Cov}:20\% $, respectively. (d) Schematic diagram of the correlation between adjacent pixels. (e) The corresponding variation of CRLB with the number of measurements under different constraints.}
	\label{constraint_space}
\end{figure}

The previous section's work was based on Eq.~(\ref{Eq_p0}), which assumed all pixels of the object are statistically independent. 
However, in reality, the object image contains specific characteristics that can be incorporated as a prior in the filtering process. 
For instance, in most cases, the object is continuous and there is a certain correlation between adjacent point, as shown in Fig.~\ref{constraint_space}(d) \cite{kim2021ghost}. 
Incorporating the prior information about object, we can apply a constraint on the spatial domain.
In Eq.~(\ref{Eq_p0}), the initial covariance matrix is $ P_0=\beta E $. 
Now, we assume that each point in the image is correlated with adjacent points, and set the initial covariance matrix to $ P_0=\beta E +\gamma\beta (E,1)+\gamma\beta (E,-1)+\gamma\beta (E,n)+\gamma\beta (E,-n) $, where $(E,n)$ denotes the upward translation of matrix $ E $ by $n$ positions.
Since we converted the two-dimensional image object into one-dimensional data during the estimation process, the subsequent four items in $ P_0 $ correspond to the surrounding points of the point in the first item.
The variable $\gamma$ denotes the correlation between each point and its surrounding points in the covariance matrix. Specifically, we assign the values 0\%, 10\%, and 20\%, respectively, using `Cov' as the symbol.

Fig.~\ref{constraint_space}(a), (b), and (c) represent the cases of the correlation (Cov) between adjacent pixels is 0\%, 10\%, and 20\% of pixel autocorrelation, respectively.
The ground truth is Fig.~\ref{airplane_1} (f), and the imaging results from left to right represents the number of measurements ranging from $600$ to $4800$.
It can be seen that as Cov increases, the imaging results converge and approach the true value faster.
Similarly, we can calculate Fisher information and CRLB during the estimation process.
Fig.~\ref{constraint_space}(e) shows the change of CRLB under different correlation constraints.
The blue solid line, orange dotted line, and yellow dotted chain line represent the case of $ \mathrm{Cov}:0\% $, $ \mathrm{Cov}:10\% $, and $ \mathrm{Cov}:20\% $ respectively.
As the constraints increase, the decline of the CRLB slope becomes steeper, indicating a faster reduction in estimation error, which aligns with the trend of image quality. Ultimately, they all lead to a decrease in the CRLB and approach zero.
Incorporating prior information accelerates the acquisition of information, reduces the CRLB faster, and speed up the image restoration.

\subsection{Adaptive Coding Based on CRLB}

Given an measure indicating the degree of acquired image information, we can utilize it to achieve adaptive encoding without relying on the imaging results. 
In this section, we introduce an adaptive encoding technique that leverages CRLB to enable rapid and efficient image reconstruction.
Object images in nature typically exhibit a concentration of information in low frequencies. 
\begin{figure}[htbp]
	\centering
	\includegraphics[width=0.9\linewidth]{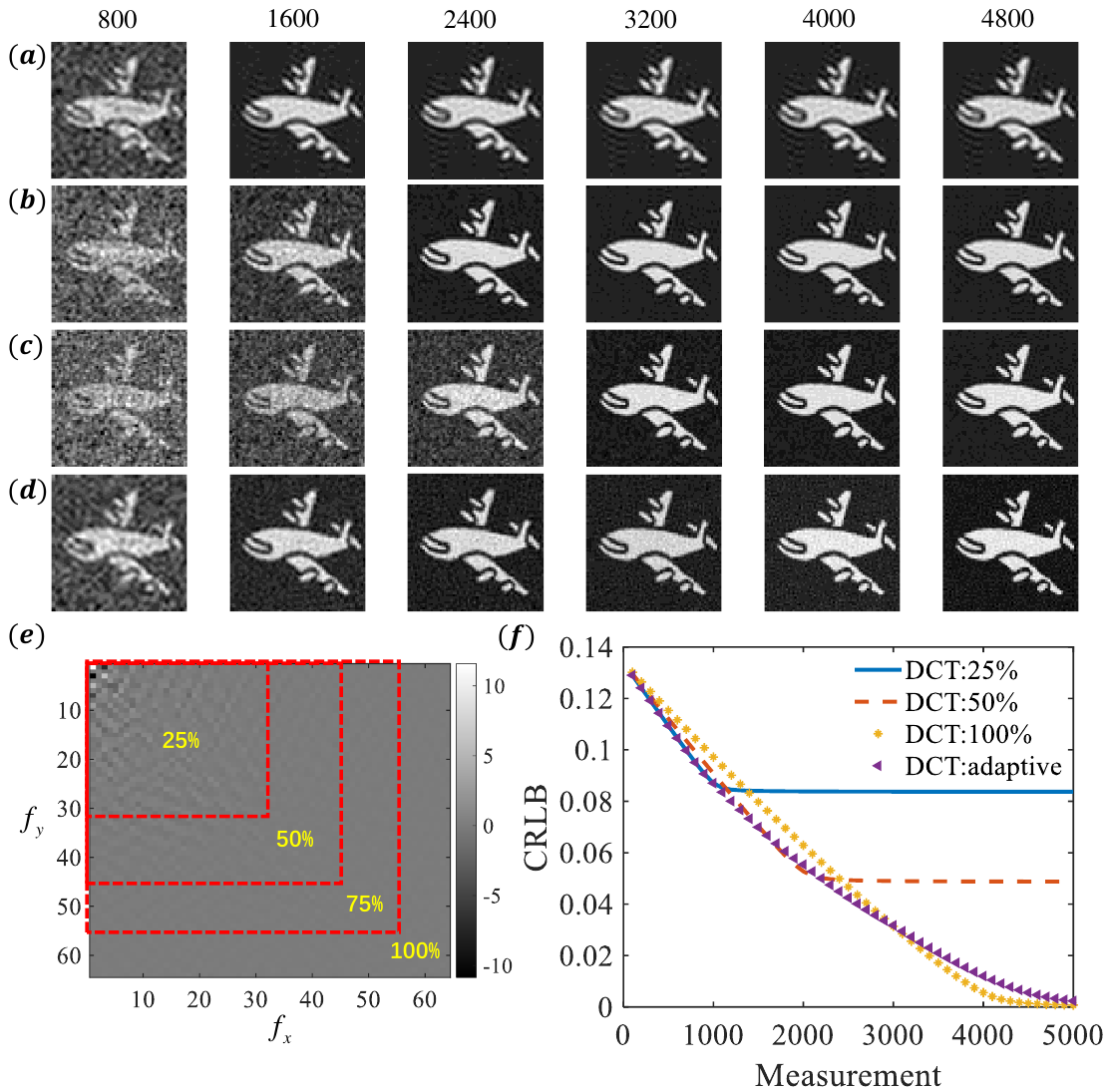}
	\caption{Adaptive coding in the frequency domain Based on CRLB. (a), (b), and (c) represent the imaging results under constraints of $ 25\% $, $ 50\% $, and $ 100\% $ in the DCT domain, respectively. (d) Imaging results of adaptive sampling for unknown scenes. (e) DCT domain representation of the object in Fig.~\ref{airplane_1}(f). (f) The corrfesponding CRLB varies with the number of measurements, under different frequency domain constraints and the adaptive sampling. }
	\label{constraint_DCT}
\end{figure}
For instance, the Discrete Cosine Transform (DCT) of the object in Fig.~\ref{airplane_1}(e) is depicted in Fig.~\ref{constraint_DCT}(e). 
The horizontal and vertical axes in Fig.~\ref{constraint_DCT}(e) represent the coordinates of the two-dimensional frequency domain of the image, revealing a concentration of larger values in the lower frequency domain.
This characteristic facilitates the implementation of "compressed sensing", which prioritizes measuring low frequencies while disregarding high frequencies \cite{donoho2006compressed}. 
Here, we adopt adaptive measurement based on CRLB in the frequency domain.
Without loss of generality, we applied the DCT transformation to the sampled speckle field and subsequently truncated it in the frequency domain. 
Specifically, as illustrated by the red dashed box in Fig.~\ref{constraint_DCT}(e), the object is encoded using frequency-limited light fields, with the truncation ranges of $ 25\% $, $ 50\% $, $ 75\% $, and $ 100\% $ in the $\mathrm{DCT}$ domain respectively. 
Incorporating the $ \mathrm{Cov}:20\% $ spatial constraints, the variations in CRLB under different frequency constraints are illustrated in Fig.~\ref{constraint_DCT}(f).
The yellow asterisk represents frequency truncation of $ \mathrm{DCT}:100\% $, which means no frequency domain constraints.
The blue solid line and red dashed line represent the changes in CRLB under frequency domain truncation $ \mathrm{DCT}:25\% $, and $ \mathrm{DCT}:50\% $, respectively.
According to the solid blue line, the CRLB decreases the fastest initially, yet asymptotically approaches a constant value, preventing it from reaching zero and consequently, limiting the imaging accuracy.
With the frequency constraint range expanded to $ \mathrm{DCT}:50\% $, the CRLB further decreases, but the initial rate of reduction has decelerated, as indicated by the red dashed line. 
Fig.~\ref{constraint_DCT} (a), (b), and (c) represent the imaging results of $\mathrm{DCT}:25\%$, $\mathrm{DCT}:50\%$, and $\mathrm{DCT}:100\%$, respectively. 
The imaging frames from left to right are $600-4800$.
The image results are consistent with the CRLB in Fig.~\ref{constraint_DCT}(f). 
The narrower the range of frequency domain constraints, the quicker the image emerges, with the most obvious manifestation in Fig.~\ref{constraint_DCT} (a). 
However, as Fig.~\ref{constraint_DCT} (a), only low-frequency information is measured and high-frequency information is missing, resulting in noticeable artifacts in the final image.
The images in Fig.~\ref{constraint_DCT}(b) and (c) appear later, but the final resolution is higher.

By estimating CRLB simultaneously, we found that with frequency domain constraints, the decline rate of CRLB is accelerated at first, but the CRLB no longer decreases after a certain number of measurements. 
In many scenarios, there is a need to acquire image information rapidly, and without interrupting the information acquisition process after a certain number of measurements.
To address this, we designed an adaptive sampling based on CRLB. We propose expanding the spatial frequency range and continuing the measurement process when the CRLB no longer demonstrates further decline. 
Specifically, a constraint of $ \mathrm{DCT}:25\% $ is applied first. When CRLB no longer decreases, the constraint is increased to $ \mathrm{DCT}:50\% $, $ \mathrm{DCT}:75\% $, and finally, samplings are taken at the full frequency range.
The purple triangle line in Fig.~\ref{constraint_DCT}(f) depicts the variations in CRLB under this imaging strategy.
The CRLB maintains a steep decrease initially and ultimately converges to zero.
The corresponding imaging results are presented in Fig.~\ref{constraint_DCT}(d), exhibiting fast imaging reconstruction at first and high accuracy in the final reconstructed imaging result, eliminating artifacts.
Meanwhile, the appearance of image is more in line with the visual requirements of the human eye \cite{phillips2017adaptive}.
This method enables adaptive design and adjustment of the imaging process independent of imaging results, which is of significant importance for CI system design. 
Furthermore, we have only presented a basic demonstration here, and there exists potential for implementing multiple adaptive methods based on this framework.

\section{EXPERIMENTAL RESULTS}

\begin{figure}[htbp]
	\centering
	\includegraphics[width=0.7\linewidth]{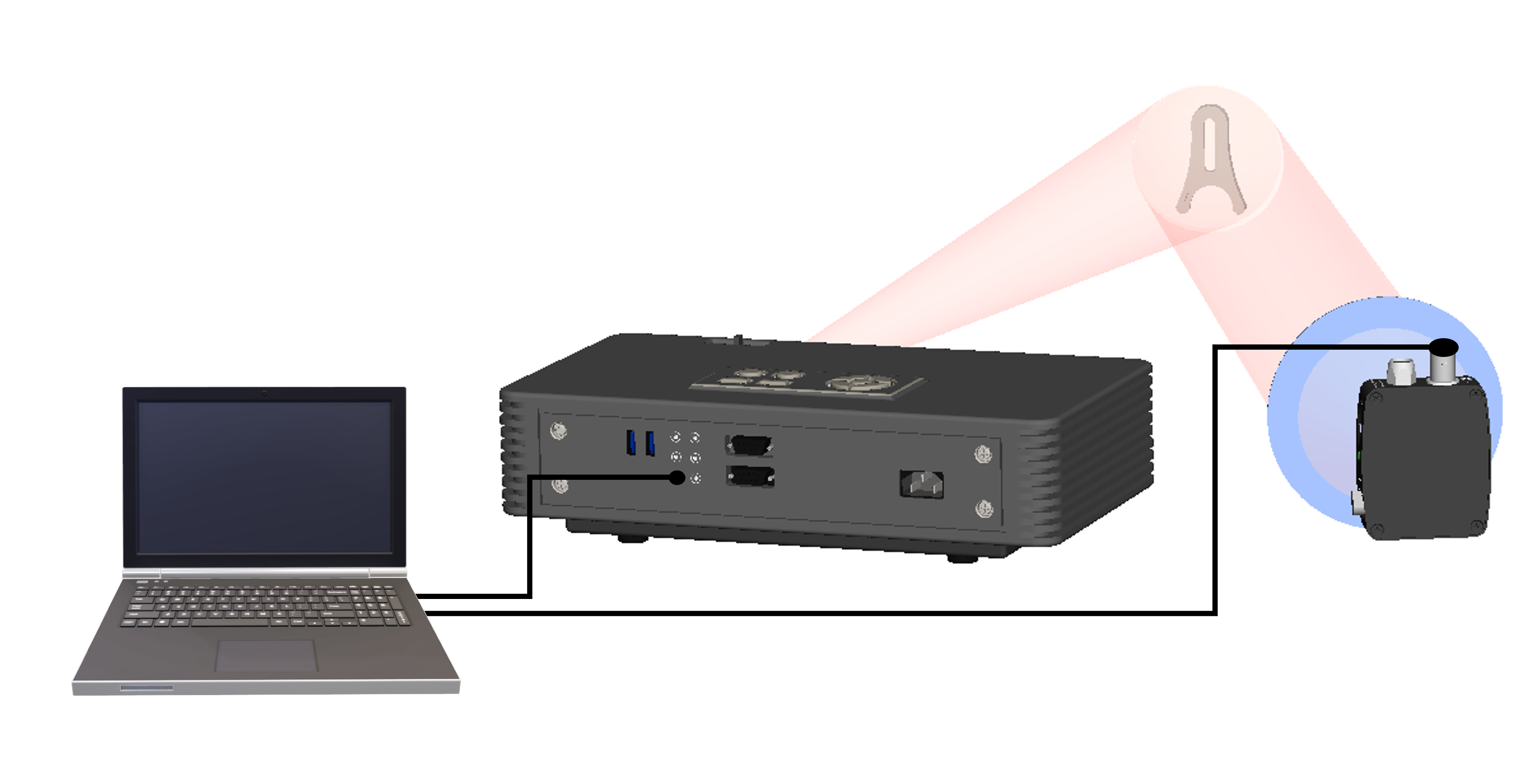}
	\caption{The experimental setup for BGI. A projector (Panasonic: PT-X301) is employed to project a pseudothermal speckle, and the echo of the object is received by a bucket detection (Thorlabs: PDA100A2). Real-time calculations enable the acquisition of image in real-time.}
	\label{experiment_setup}
\end{figure}

To further demonstrate the ability of our methods, we conducted an experimental demonstration of BGI, as shown in Fig.~\ref{experiment_setup}.

Pseudothermal light field are generated via a computer-controlled projector, and real-time echo data is acquired through bucket detection.
In the experiment, we calibrated the brightness of the projector and projected an illumination field with a negative exponential distribution ranging from 0 to 1.
The field of view consists of 40 * 40 pixels, with each pixel being statistically independent.
The initial probability density of the object is the same as Eq.~(\ref{Eq_p0}).
\begin{figure}[htbp]
	\centering
	\includegraphics[width=0.9\linewidth]{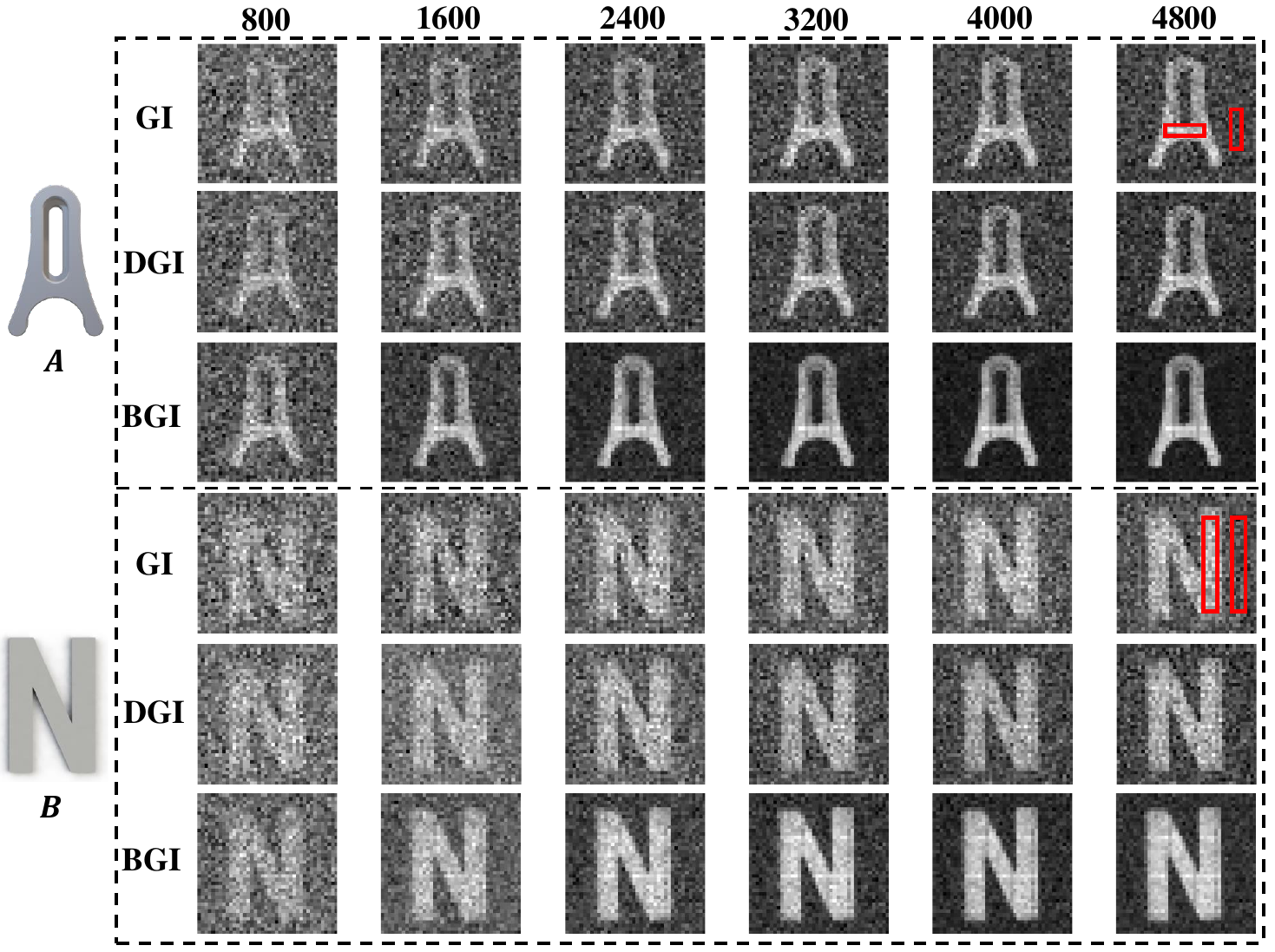}
	\caption{
	The experimental imaging results of a clamping Fork and a 3D printed letter, denoted as `A' and `B'. The upper three lines on the right represent the GI, DGI, and BGI results of `A', while the lower three lines represent the GI, DGI, and BGI results of `B'.}
	\label{experiment_results_1}
\end{figure}
The imaging results of BGI are calculated in real time and the specific experimental results are shown in Fig.~\ref{experiment_results_1}.
The object `A' is a clamping Fork (DHC:GCM-5328M), a component of the experiment. Object `B' is a 3D printed letter 'N'.
The upper three rows represent the imaging results of Object A, corresponding to GI, DGI, and BGI, respectively. Similarly, the lower three rows represent the imaging results of Object B, corresponding to GI, DGI, and BGI as well.
The number of samples from left to right is 800-4800. 
It is evident from Fig.~\ref{experiment_results_1} that, for both object A and object B,  the results of DGI are slightly better than those of GI, and the BGI results demonstrate a significant improvement over both the GI and DGI results, particularly with a greater number of measurements.

\begin{figure}[htbp]
	\centering
	\includegraphics[width=0.8\linewidth]{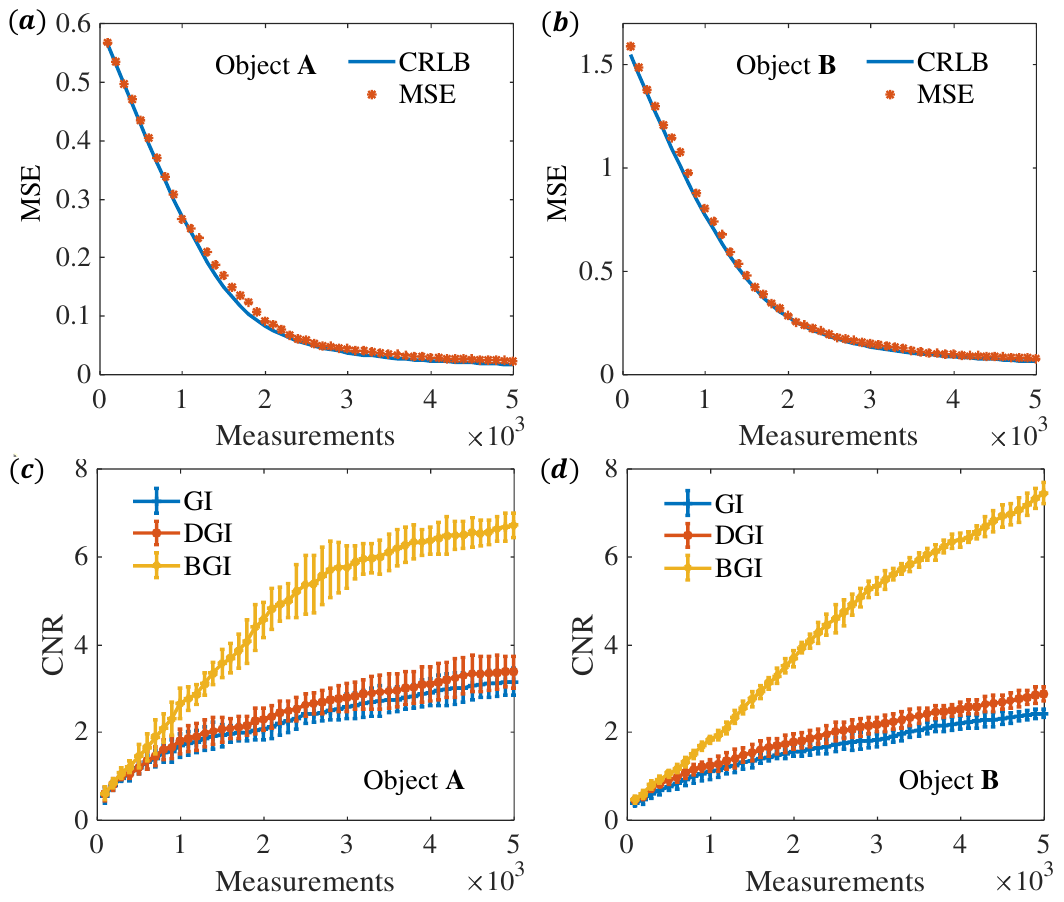}
	\caption{ The analysis and comparison of experimental imaging results. (a) and (b) are the changes in CRLB and MSE of objects A and B in BGI, with respect to the number of measurements. The predicted CRLB is represented by the solid blue line, and the MSE of image is denoted by the red asterisk. (c) and (d) represent the imaging results of objects A and B with GI, DGI, and BGI. The solid blue, red, and yellow lines represent the imaging quality results of GI, DGI, and BGI, respectively. }
	\label{experiment_results_2}
\end{figure}

In the experiment, CRLB was predicted in advance, and the MSE of the imaging results was calculated during the imaging process.
As mentioned earlier, the initial $t_0$ and $P_0$ was determined by the bucket detection signals.
Specifically, for object A, they are $t_0=0.56 \boldsymbol{e}$ and $P_0=0.60 E$, and for object B, they are $t_0=1.24 \boldsymbol{e}$ and $P_0=1.64 E$.
Subsequently, for various objects, the noise level of the detector $ R $ was measured under constant illumination condition.
Under these conditions, the change in CRLB can be calculated in advance based on the projected speckle field.
The variations in CRLB during the measurement process of object A and B are shown in Fig.~\ref{experiment_results_2}(a) and Fig.~\ref{experiment_results_2}(b), respectively, represented by the solid blue lines.
Additionally, the MSE in Eq.~(\ref{Eq_mse}) of the imaging results are calculated during the measurement process.
To calculate the MSE, the true image of the object is obtained through point scanning.
The results of MSE is indicated  by the red asterisk in Fig.~\ref{experiment_results_2}(a) and Fig.~\ref{experiment_results_2}(b). 
Notably, the trend of MSE closely aligns with CRLB, thereby validating the efficacy of our method.

Moreover, we conducted a quantitative comparison of the imaging quality in Fig.~\ref{experiment_results_2} (c) and (d). For the convenience of comparing GI, DGI, and BGI simultaneously, Contrast-to-Noise Ratio (CNR) is used as a measure, as Eq.~(\ref{eqs_CNR}) \cite{chan2009high}.
\begin{equation} 
	G_{\mathrm {CNR}} = \frac{{{{\left\langle  G(\bm{\rho}_{in})\right\rangle }} - {{\left\langle G(\bm{\rho}_{out}) \right\rangle }}}}{\sqrt {\Delta^2 G(\bm{\rho}_{in}) + \Delta^2 G(\bm{\rho}_{out}) }},
	\label{eqs_CNR}
\end{equation}
where $ \bm{\rho}_{in} $ represents the area occupied by the object, called object region, and $ \bm{\rho}_{out} $ represents the rest of the area, called background region. 
$G$ represents the result of the image and $ \Delta^2 G $ is the corresponding variance.
The normalization of GI and DGI in the experiment has no impact on the results of CNR.
The objects are approximated as binary objects, and the areas within the red box in Fig.~\ref{experiment_results_1} are selected as the object regions and background regions. For object A and object B, the selected area sizes are 3*16 and 4*24, respectively.
The solid blue, red, and yellow lines in Fig.~\ref{experiment_results_2} (c) and (d) correspond to the imaging quality results of GI, DGI, and BGI, respectively.
Each data point is derived from 10 sets of experimental data, and the mean and errorbar are provided.
Consistent with the image results in Fig.~\ref{experiment_results_1}, the performance of BGI surpasses that of GI and DGI.


\section{CONCLUSION AND DISCUSSION}

To summarize, in this paper, a quantitative framework for CI systems is proposed where the Bayesian filtering paradigm is incorporated to give a reliable evaluation on the intrinsic ability of acquiring information and the imaging performance bound. The proposed scheme is demonstrated based on the typical GI system. 
The method for modeling GI system as a Bayesian-filtering problem is provided, and the scheme for recursively estimating the object's image information and its uncertainty is developed and demonstrated. In specific, the Fisher information matrix is exploited to calculate the recursive estimation of the acquired information, and CRLB is quantified for the object image retrieval.
Through numerical simulations and practical experiments, we validate the capability and superiority of the proposed scheme on three aspects, 1) the ability on quantitatively estimating the acquired information and the resulting imaging performance bound, 2) the enhanced quality of retrieved image information compared with those linear estimation algorithms, and its consistency with the information-theoretic CRLB, and 3) the scalability on incorporating various kinds of image priors to achieve quality enhancement and adaptive encoding design. 
In a nutshell, the proposed scheme enables not only the quantitative performance evaluation, but also the adaptive encoding design for performance optimization, and is thus expected to provide a reliable approach for information-quantitative CI techniques and applications.

Furthermore, we have performed a basic adaptive coding technique based on CRLB in this article, and there are numerous possibilities for exploring applications that leverage the information.
For moving objects, it is worth noting that the motion information and image information can be considered as parallel and decoupled~\cite{du2023information}. 
Hence, it is supposed to image moving objects by incorporating their motion state information into Bayesian filtering models. 
Therefore, Bayesian filtering can be employed to simultaneously track and gradually image moving object, which is the focus of our future work.

\appendix
\section*{Appendix: Setting of initial filtering values in BGI.} 

In BGI, object information can be estimated from bucket detection signals, allowing the initial value of BGI to be determined based on this information. The specific method is as follows.
The illumination light field is pseudothermal with a negative exponential distribution. 
According to Eq.~(\ref{Eq_Ib}) to Eq.~(\ref{Eq_Ir}), bucket detection can be written as ${I_b} = \int k {I({\boldsymbol{\rho}_r})} t({\boldsymbol{\rho}_r})d{\boldsymbol{\rho}_r}$, where $ k $ is a constant.
Without loss of generality, $ k $ can be considered equal to one and the bucket detection can be written as
\begin{equation}
	{I_b} = \int {I({\boldsymbol{\rho}_r})} t({\boldsymbol{\rho}_r})d{\boldsymbol{\rho}_r},
\end{equation}
The mean of the object can be expressed as
\begin{equation}
	\bar{t} = \frac {\int t({\boldsymbol{\rho}_r}) \mathrm{d} {\boldsymbol{\rho}_r}} {A_\mathrm{o}} = \frac {\left\langle I_b \right\rangle}{ \left\langle I_r \right\rangle} \cdot \frac{1}{A_{\mathrm{o}}},
	\label{t_bar}
\end{equation}
where $\left\langle I_r \right\rangle$ represents the mean of the illumination light field, $\left\langle I_b \right\rangle$ represents the mean of bucket detection, and $A_\mathrm{o}$ is the size of the object area of interest.
Besides, the mean square error of the object is
\begin{equation}
	t_\mathrm{MSE}=\frac{\int[t(\boldsymbol{\rho}_r)-\bar{t}]^2 \mathrm{~d} \boldsymbol{\rho}_r}{A_{\mathrm{o}}}=\frac{\int t^2(\boldsymbol{\rho}_r) \mathrm{d} \boldsymbol{\rho}_r}{A_{\mathrm{o}}}-\bar{t}^2.
	\label{t_mse}
\end{equation}
According to the circular complex Gaussian characteristics of the pseudothermal light field \cite{goodman2015statistical}, it can be obtained that 
\begin{equation}
	\left\langle I_{b}^2\right\rangle =\left\langle I_{r}\right\rangle^2\left[\int t(\boldsymbol{\rho}_r) \mathrm{d} \boldsymbol{\rho}_r \right]^2+\left\langle I_{r}\right\rangle^2 A_\mathrm{coh} \int t^2(\boldsymbol{\rho}_r) d \boldsymbol{\rho}_r,
	\label{I_b2}
\end{equation}
where $A_\mathrm{coh}$ represents the coherent area of the illumination light field, i.e., the size of the speckle.
Substituting Eq.~(\ref{t_bar}) and Eq.~(\ref{I_b2}) into Eq.~(\ref{t_mse}) yields that
\begin{equation}
	t_\mathrm{MSE}=\frac{\left\langle I_b^2\right\rangle-\left\langle I_b \right\rangle^2}{\left\langle I_r\right\rangle^2  A_{\mathrm{o}} A_\mathrm{coh}}-\frac{\left\langle I_b \right\rangle^2}{\left\langle I_r \right\rangle^2} \cdot \frac{1}{A_{\mathrm{o}}^2}.
\end{equation}
Therefore, we can estimate the mean and variance of objects based on the bucket detection.

\begin{backmatter}
	
\bmsection{Funding} 
National Natural Science Foundation of China (62105365, 62275270, 62201165); Science Fund for Distinguished Young Scholars of Hunan Province (2021JJ10051); Research Program of National University of Defense Technology (ZK22-58);

\bmsection{Disclosures}
The authors declare no conflicts of interest.

\bmsection{Data availability} Data underlying the results presented in this paper are not publicly available at this time but may be obtained from the authors upon reasonable request.

\end{backmatter}

\bibliography{Bayes}

\end{document}